\documentclass[aip,jcp,reprint]{revtex4-1} 
\usepackage{graphicx}
\usepackage{enumitem}
\usepackage{float}

\usepackage{mathtools}
\usepackage{graphicx}   
\usepackage{color}
\usepackage{ulem}
\usepackage{caption}
\usepackage{subcaption}
\usepackage{comment}
\usepackage{bm}

\usepackage{array}
\usepackage{booktabs}
\usepackage{multirow}

\usepackage{amsmath}
\newcommand\numberthis{\addtocounter{equation}{1}\tag{\theequation}}

\usepackage{booktabs}

\begin{document}
\title{Understanding the role of predictive time delay and biased propagator in RAVE}

	\author{Yihang Wang}
 \affiliation{Biophysics Program and Institute for Physical Science and Technology,
 University of Maryland, College Park 20742, USA.}

 \author{Pratyush Tiwary*}
 \affiliation{Department of Chemistry and Biochemistry and Institute for Physical Science and Technology,
 University of Maryland, College Park 20742, USA.}

	\date{\today}
	
	\begin{abstract}
In this work, we revisit our recent iterative machine learning (ML) -- molecular dynamics (MD) technique ``Reweighted autoencoded variational Bayes for enhanced sampling (RAVE)" (Ribeiro, Bravo, Wang, Tiwary, J. Chem. Phys. \textbf{149} 072301 (2018) and Wang, Ribeiro, Tiwary, Nature Commun. \textbf{10} 3573 (2019)) and analyze as well as formalize some of its approximations. These including: (a) the choice of a predictive time-delay, or how far into the future should the ML try to predict the state of a given system output from MD, and (b) for short time-delays, how much of an error is made in approximating the biased propagator for the dynamics as the unbiased propagator. We demonstrate through a master equation framework as to why the exact choice of time-delay is irrelevant as long as a small non-zero value is adopted. We also derive a correction to reweight the biased propagator, and somewhat to our dissatisfaction but also to our reassurance, find that it barely makes a difference to the intuitive picture we had previously derived and used. 
\end{abstract}

	\maketitle
\section{Introduction}
We recently proposed a mixed machine learning (ML) --molecular dynamics (MD) scheme that by systematically iterating between rounds of ML and MD, makes it possible to explore configuration space of complex molecules in an enhanced manner, yet efficiently obtain equilibrium estimates of thermodynamic and kinetic observables. The method and its subsequent variants are collectively named ``Reweighted Autoencoded Variational Bayes for Enhanced Sampling (RAVE)".\cite{rave,multirave,prave,cosb_ml} The first central idea in RAVE is that ML or more specifically deep learning, used here in a form akin to the deep variational information bottleneck framework\cite{dvib}, can aid in recovering the  important low-dimensional features of a given MD trajectory.  These features are collectively called the ``past--future information bottleneck" (PIB). Per construction, the PIB is the minimally complex yet at the same time maximally predictive representation of the trajectory's evolution slightly ahead in future. RAVE considers the PIB as a computationally tractable proxy for the reaction coordinate (RC) that is ubiquitous in theoretical and computational chemistry, yet hard to calculate. The second central idea is that the probability distribution of the PIB can be used as an importance function (or bias) in the next round of MD based sampling, encouraging the MD to explore regions of the configuration space which would have otherwise been rarely visited. This improved exploration, appropriately reweighted, in turn leads to a better estimate of the PIB and its probability distribution, which then amounts to a better sampling bias for further systematic and enhanced exploration. Thus, solving the equations of motion of the system generates data which when used in a learning algorithm identifies efficient ways to speed-up the dynamics, which in turn generates more data for improved learning, and so on. This iterative algorithm is terminated when the PIB and the associated bias do not change with further rounds, and their use is sufficient for back-and-forth transitions between various metastable states. Such a static bias is then used in routine MD simulations which run orders of magnitude faster than unbiased MD, yet, through the use of reweighting they allow obtaining equilibrium probability and kinetic rates even from biased trajectories. This protocol was demonstrated to be useful and accurate through various applications. \cite{rave,multirave,prave}

In this work, we take a closer look at the underlying formulation of RAVE with respect to crucial aspects which were introduced in our past work using intuitive arguments. Here we put these aspects on a rigorous footing, identifying their shortcomings and when applicable introducing simple corrections. These are as follows. RAVE\cite{prave} involves learning the PIB through optimization of an objective function, detailed in Sec. \ref{sec:rave_summary}, that minimizes model complexity while still maximizing predictive power. Firstly, a key parameter in constructing the RAVE objective function is the so-called predictive time-delay $\Delta t$, or how far into the future should the algorithm try to predict. In Ref. \cite{prave} we had shown through numerical benchmarks that the precise choice of $\Delta t$ did not matter, as long as $\Delta t >0$ and was kept small. That is, the predicted PIB changed but rapidly plateaued with increase in $\Delta t$. Here, by using a Master Equation based formulation of the system's dynamics, we demonstrate rigorously why such a dependence on $\Delta t$ is a direct and simple consequence of Markovianity along the learnt low-dimensional representation. In fact, the absence of a plateau in how the PIB depends on $\Delta t$ can even be considered a tell-tale sign of incorrect results from RAVE.

Secondly, at the initial stage of RAVE, when the input trajectory is unbiased, the RAVE objective function can be computed exactly by sampling over the input trajectory. However, after the first iteration once the PIB is learnt and used to perform a round of biased MD, it becomes tricky to calculate the objective function exactly through sampling.  Here as well Ref. \cite{prave} introduced an intuitive approximation, namely that at short timescales the propagator of the system along the putative PIB is invariant between biased and unbiased dynamics. Naturally this statement is exactly true as $\Delta t \mapsto 0$, but its fate for $\Delta t>0$ is not at all obvious. Here, again by taking recourse in a Master equation formalism, we derive a framework for using biased sampling to estimate a reweighted, unbiased propagator needed in RAVE that is valid for small $\Delta t >0$. We show how it reduces to the original expression in limiting cases, and how to compute it. We also demonstrate through numerical examples when such a correction might matter and when it might not. We thus believe this work presents a useful theoretical investigation into the underpinnings of RAVE, so the community can use it with greater confidence for the simulation of complex molecules plagued with rare events.

\section{theory}
\label{sec:theory}

\subsection{Summarizing RAVE}
\label{sec:rave_summary}

RAVE is proposed as a framework which iterates between constructing the the reaction coordinate (RC) through the use of deep learning, and performing biased MD making use of this RC and its associated thermodynamic/kinetic information learnt in deep learning. The central idea in RAVE is to interpret the RC as a predictive information bottleneck (PIB), which means that it should be a minimally complex but maximally predictive representation of the high-dimensional trajectory. Mathematically, finding the PIB can be formulated as maximizing the the objective function:
\begin{equation} 
\label{eq:IB_obj}
\mathcal{L}\equiv I(\bm{\chi},\mathbf{X}_{\Delta t})  - \gamma I(\mathbf{X},\bm{\chi})
\end{equation} 
Here $\bm{\chi}$ is the low-dimensional PIB, $\mathbf{X}$ and $\mathbf{X}_{\Delta t}$ are high-dimensional basis functions or order parameters that characterize the state of the system at two instances of time separated by a time delay $\Delta t$, and $I(\mathbf{X},\mathbf{Y})$ denotes the mutual information between any two random variables $\mathbf{X}$ and $\mathbf{Y}$. As reducing the mutual information between $\mathbf{X}$ and $\bm{\chi}$ will limit the ability of our model to predict $\mathbf{X}_{\Delta t}$, there is a competition between compressing information and predicting the future. This competition is controlled through a model complexity parameter $\gamma$, which determines the trade-off between compression of the information from $\mathbf{X}$ and the ability to predict $\mathbf{X}_{\Delta t}$. Maximizing $\mathcal{L}$ is equivalent to maximizing $I( \bm{\chi} ,\mathbf{X}_{\Delta t})$ and minimizing $I(\mathbf{X},\bm{\chi})$ simultaneously. By minimizing $I(\mathbf{X},\bm{\chi})$, we find a mapping $P(\bm{\chi}|\mathbf{X})$ which gives a representation $\bm{\chi}$ carrying as little information about the past $\mathbf{X}$ as possible. Given $P(\bm{\chi}|\mathbf{X})$ and $P(\mathbf{X}, \mathbf{X}_{\Delta t})$, the mapping $P(\bm{\mathbf{X}_{\Delta t}|\chi})$  can be determined either exactly through Bayes' theorem, or more practically but approximately by variational inference.\cite{prave} $P(\bm{\mathbf{X}_{\Delta t}|\chi})$ can be used to calculate  $I(\mathbf{X}_{\Delta t}, \bm{\chi}$). The mappings $P(\bm{\chi}|\mathbf{X})$  and $P(\bm{\mathbf{X}_{\Delta t}|\chi})$  can be stochastic which helps with avoiding overfitting among other benefits, and are generally refered to as encoder and decoder mappings respectively.

In a recent work\cite{prave}, we developed a protocol for learning such a PIB wherein we proposed the use of an interpretable linear encoder  $P(\bm{\chi}|\mathbf{X})$, and a stochastic deep neural network decoder $P(\bm{\mathbf{X}_{\Delta t}|\chi})$. In principle, we can also use a non-linear encoder but a linear encoder gives us a directly physically interpretable PIB, while avoiding optimization issues inherent in non-linear functions. Furthermore, given the limited complexity that a linear encoder can achieve, we set in Eq.\ref{eq:IB_obj} the hyper-parameter $\gamma=0$. If we define the PIB $\chi = \sum_i w_i X_i$, the values of $w_i$ reflect the relative importance of order parameters $X_i$. Running estimate of the free energy along $\bm{\chi}$ is used to construct a biasing potential to perform biased MD simulation. If this PIB is indeed the RC i.e. it captures relevant slow modes and any missing modes are either fast or irrelevant, this biased simulation should lead to increased sampling. Such an improved sampling can then be re-fed to the encoder-decoder architecture for learning an improved PIB or RC. The iteration can then be continued until ergodic sampling is achieved.  



\subsection{Dependence of PIB on $\Delta t$}
Maximizing the mutual information $I(\bm{\chi},\mathbf{X}_{\Delta t})$ in Eq. \ref{eq:IB_obj} forces the learnt linear PIB or RC $\chi$ to be as predictive as possible of the high probability aspects of the feature space $\mathbf{X}_{\Delta t}$. In addition, introducing a slight time delay in RAVE i.e. setting $\Delta t >0$ emphasizes the contribution of not just high probability aspects of the data, but specifically of high probability aspects that persist with time. As demonstrated for related work on time-lagged autoencoders\cite{time-lagged_ae}, this amounts to learning slowly decorrelating aspects of the feature space $\mathbf{X}$.

 Two questions immediately arise after introducing such a non-zero time delay $\Delta t$. First, what precise value of $\Delta t$ should we adopt when constructing the objective function? Second, how to approximate unbiased estimates of $P(\mathbf{X}_{\Delta t}  | \mathbf{X})$ from a biased MD trajectory which provides us $P_\text{biased}(\mathbf{X}_{\Delta t}  | \mathbf{X})$? The second question arises because access to unbiased estimates of $P(\mathbf{X}_{\Delta t}  | \mathbf{X})$ is critical to calculating the objective function $\mathcal{L}$ in Eq. \ref{eq:IB_obj}.

 In this work we answer, with relative rigor, both the questions above. For the first question, we demonstrate that given markovianity in the higher dimensional space $\mathbf{X}$, the markovianity of the PIB follows quite naturally. In other words, there exists a range of non-zero time-delay $\Delta t$ values using which the PIB is independent of the exact choice of $\Delta t$. Thus any small enough $\Delta t$ can be used to construct the PIB. For the second question, we propose a new objective function which corrects for the influence of bias on the short-time propagator $P(\mathbf{X}_{\Delta t}|\bm{\chi})$, and test it on model systems. Reassuringly we find that the effect of the correction derived here is minimal, and our intuitive approximation of $P_\text{biased}(\mathbf{X}_{\Delta t}|\bm{\chi}) \approx  P_\text{unbiased}(\mathbf{X}_{\Delta t}|\bm{\chi}) $ made in Ref. \onlinecite{prave} was not a bad one. 

In Ref.\cite{prave}, we showed that estimating the PIB, or minimizing the objective function $\mathcal{L}$ in Eq. \ref{eq:IB_obj} is the same as maximizing the cross entropy $\mathcal{L}'$ between $P_{\theta}(\mathbf{X}_{\Delta t},  \bm{\chi})$ and $P_{\theta}( \mathbf{X}_{\Delta t } | \bm{\chi} )$: 
\begin{equation}
\mathcal{L}' = -\int P_{\theta}(\mathbf{X}_{\Delta t},  \bm{\chi}) \ln P_{\theta}( \mathbf{X}_{\Delta t } | \bm{\chi} )d\mathbf{X}_{\Delta t }d\bm{\chi}  
\label{eq:lprime}
\end{equation}
where $\theta$ indicates the parameters of neural network. To understand how this objective function $\mathcal{L}'$ depends on the predictive time delay $\Delta t$, we need to analyze the $\Delta t$-dependence of $P_{\theta }( \mathbf{X}_{\Delta t} |  \bm{\chi})$ and $P_{\theta}( \mathbf{X}_{\Delta t},  \bm{\chi})$. To do so we start with a master equation framework for the propagator in the high-dimensional space $\mathbf{X}$, where we assume markovianity holds true:
\begin{align*}
\label{eq:mastereq1}
 P(\mathbf{X}_2,t+\Delta t | \mathbf{X}_1, t)& = \delta(\mathbf{X}_2 -\mathbf{X}_1)[1-a(\mathbf{X}_1,t)\Delta t]  
\\ &+ W_t(\mathbf{X}_2|\mathbf{X}_1)\Delta t + O[(\Delta t)^2] \numberthis
\end{align*}
Here $W_t(\mathbf{X}_2|\mathbf{X}_1)$ is the transition probability per unit time from $\mathbf{X}_1$ to $\mathbf{X}_2$ at time $t$, and $a(\mathbf{X}_1,t)=\int W_t(\mathbf{X}_2|\mathbf{X}_1)d\mathbf{X}_2$. 

At this point we make two assumptions: (a) the distribution is stationary so $ P(\mathbf{X}_2,t+\Delta t | \mathbf{X}_1, t)$ can be denoted as $P(\mathbf{X}_{\Delta t} | \mathbf{X})$, and (b) $\Delta t$ is small enough so higher order terms in Eq. \ref{eq:mastereq1} can be ignored giving:
\begin{align*}
\label{eq:mastereq2}
 P(\mathbf{X}_{\Delta t}| \mathbf{X}) = \delta(\mathbf{X}_{\Delta t} -\mathbf{X})[1-a(\mathbf{X})\Delta t] + W(\mathbf{X}_{\Delta t}|\mathbf{X})\Delta t 
 \numberthis
\end{align*}

In addition, as a direct consequence of the markovianity assumption in $\mathbf{X}$ space, the following property must hold true for the models we discuss \cite{PIB,dvib}:
\begin{align*}
\label{Markovian}
 P(\mathbf{X}_{\Delta t}| \bm{\chi}, \mathbf{X}) =  P(\mathbf{X}_{\Delta t}|  \mathbf{X})  \numberthis
\end{align*}

This property reflects that $\mathbf{X}$ contains all the information needed to predict $\mathbf{X}_{\Delta t}$, and in addition using knowledge of $\bm{\chi}$ can not improve the quality of prediction. With this additional assumption and the use of Eq. \ref{eq:mastereq2}, we have:
\begin{align*}
 P\left(\mathbf{X}_{\Delta t}, \bm{\chi}\right) = & \int P\left(\bm{\chi},\mathbf{X}\right) P\left(\mathbf{X}_{\Delta t}|\mathbf{\chi},\mathbf{X}\right)d\mathbf{X}  \\
=  & \text{ }P(\bm{\chi},\mathbf{X}_{\Delta t}) [1-a(\mathbf{X}_{\Delta t})\Delta t ]  \\
&+\Delta t  \int P(\bm{\chi},\mathbf{X})   W(\mathbf{X}_{\Delta t}|\mathbf{X}) d\mathbf{X} \numberthis \label{eq:expand_p}
\end{align*}

By rearranging Eq. \ref{eq:expand_p}, we can express $P(\mathbf{X}_{\Delta t}, \bm{\chi})$ in terms of $W(\mathbf{X}_{\Delta t}|\mathbf{X})$ and $P(\bm{\chi},\mathbf{X}) $:
\begin{align*}
P(\mathbf{X}_{\Delta t}, \bm{\chi})& = \frac{ \int P(\bm{\chi},\mathbf{X})   W(\mathbf{X}_{\Delta t}|\mathbf{X}) d\mathbf{X} }{a(\mathbf{X}_{\Delta t})}\\
& = \frac{ \int P(\bm{\chi},\mathbf{X})   W(\mathbf{X}_{\Delta t}|\mathbf{X}) d\mathbf{X} }{ \int  W(\mathbf{X}_{\Delta t'}|\mathbf{X}_{\Delta t}) d\mathbf{X}_{\Delta t'} } \numberthis \label{eq:jointP}
\end{align*}

 In Eq. \ref{eq:jointP} both the numerator and denominator have any dependence on the predictive time-delay $\Delta t$ only through the transition probability per unit time $W$, which in turn per construction does not depend on $\Delta t$. As such, $P(\mathbf{X}_{\Delta t},\chi)$ does not depend on $\Delta t$. As a direct consequence of this, the PIB estimated by maximizing $\mathcal{L}'$ in Eq. \ref{eq:lprime} as well should not depend on the choice of $\Delta t$, as long as $\Delta t$ is small enough that the Master Equation formalism of Eq. \ref{eq:mastereq2} is valid.

\subsection{Correcting the propagator and the information bottleneck when using biased trajectory}
So far the formalism assumes we have access to various unbiased estimates needed in order to maximize  $\mathcal{L}'$ in Eq. \ref{eq:lprime}. However, the whole framework in RAVE is based on iterations between machine learning and progressively biased MD, we need to be able to construct PIB from biased trajectories as well. We now address the question of how to do so, and develop a corrected form of Eq. \ref{eq:lprime} for this purpose. In our previous work,\cite{prave} we worked with biased trajectories by making the assumption:
\begin{equation}
\label{eq:approximation}
    P_\textrm{biased}(\mathbf{X}^{n+k}|\bm{\chi}^n) \approx P_\textrm{unbiased}(\mathbf{X}^{n+k}|\bm{\chi}^n)
\end{equation}
which led to the following $\mathcal{L'}$:
\begin{equation} 
\label{eq:PIB_vb3}
\mathcal{L'} = \left\{ {\sum\limits_{n=1}^{N} e^{\beta V^n}  } \right\}^{-1} { \sum\limits_{n=1}^{M} e^{\beta V^n}   \ln Q(\mathbf{X}^{n+k}|\bm{\chi}^n)    }
\end{equation} 
Here $\Delta t$ equals the time elapsed in $k$ MD steps and $Q(\mathbf{X}^{n+k}|\bm{\chi}^n)$ is the probabilistic mapping given by the approximate decoder to approximate the real posterior probability $P(\mathbf{X}^{n+k}|\bm{\chi}^n)$. $e^{\beta V^n} $ is a reweighting factor to correct for the bias in stationary probability density estimate. The need for an approximate decoder arises from the principle of variational inference wherein $\mathcal{L'}$ is bounded from above by $\mathcal{L}$ and will reach maximum only when $Q(\mathbf{X}^{n+k}|\bm{\chi}^n) = P(\mathbf{X}^{n+k}|\bm{\chi}^n)$. This guarantees that maximizing $\mathcal{L'}$ will also force the approximate decoder be as close to the real posterior probability as possible. Thus in the limit that the neural network decoder is flexible enough to approximate the real decoder exactly, optimizing $\mathcal{L}$ and $\mathcal{L'}$ will give the same encoder. As such we can assume the decoder is exact when deriving the correction for the propagator.

Eq. \ref{eq:PIB_vb3} as introduced in Ref. \onlinecite{prave} corrects for only one of the two biased aspects of the trajectory: (i) it reweights out the effect of the bias on the sampling probability at any given moment of time, (ii) it however ignores the effect of bias on the short-term dynamical evolution of the system as it assumes Eq. \ref{eq:approximation}. Note that in the limit $\Delta t \to 0$, Eq. \ref{eq:approximation} becomes exact. But as $\Delta t$ increases, $P_\textrm{biased}(\mathbf{X}|\bm{\chi})$ gradually deviates from $P_\textrm{unbiased}(\mathbf{X}|\bm{\chi})$ due to the existence of bias.

To correct for the bias in $P_\textrm{biased}(\mathbf{X}|\bm{\chi})$, we first consider  the relationship of transition probability between biased and unbiased MD. By discretizing the Smoluchowski equation along $\mathbf{X}$, the transition probability can be written as \cite{Szabo_electron_transfer,diffusion_anisotropy}:
\begin{align*}
W(\mathbf{X}_{\Delta t}|\mathbf{X}) = \frac{D(\mathbf{X}_{\Delta t} )+D(\mathbf{X})}{2||\mathbf{X}_{\Delta t}-\mathbf{X}||^2}\exp\{ -\frac{\beta[U(\mathbf{X}_{\Delta t})-U(\mathbf{X})]}{2}\} 
\numberthis 
\end{align*}where $U(\mathbf{X})$ is the equilibrium, unbiased free energy. In biased MD with bias $V(\bm{\chi}(\mathbf{X}))$ applied as a function of a putative RC $\bm{\chi}$, the underlying free energy gets replaced by  $U(\mathbf{X})+V(\bm{\chi}(\mathbf{X}))$. Note that the bias is function of RC $\bm{\chi}$ which itself is a well-defined function of basis functions or order parameters $\mathbf{X}$ that depend on atomic coordinates, so the biasing force added on atoms is continuously differentiable. Assuming that the diffusivity itself stays unchanged due to the addition of bias (a common assumption in dynamical reweighting algorithms such as Ref. \onlinecite{rosta2015free}), we can then write down a relation between the biased and unbiased transition probabilities in terms of the bias:
\begin{align*}
\label{eq:w_correction}
W_b(\mathbf{X}_{\Delta t}|\mathbf{X})=  \sqrt{\frac{e^{\beta V(\chi(\mathbf{X}))}}{e^{\beta V(\chi(\mathbf{X}_{\Delta t}))}}} W_u(\mathbf{X}_{\Delta t}|\mathbf{X})  \numberthis 
\end{align*}
where the subscripts $u$ and $b$ denote unbiased and biased measurements respectively. Eqs. \ref{eq:mastereq2} and \ref{eq:w_correction} together provide the relationship between $P_b(\mathbf{X}_{\Delta t}| \bm{\chi})$ and $P_u(\mathbf{X}_{\Delta t}| \bm{\chi})$:
\begin{align*}
&P_b(\mathbf{X}_{\Delta t}|\mathbf{X})=  \sqrt{\frac{e^{\beta V(\chi(\mathbf{X}))}}{e^{\beta V(\chi(\mathbf{X}_{\Delta t}))}}} P_u(\mathbf{X}_{\Delta t}|\mathbf{X})  &\textrm{if }\mathbf{X}_{\Delta t} \neq \mathbf{X} \\
&P_b(\mathbf{X}_{\Delta t}|\mathbf{X})= 1- \int_{\mathbf{X}'_{\Delta t} \neq \mathbf{X}_{\Delta t} } P_b(\mathbf{X}'_{\Delta t}|\mathbf{X})d\mathbf{X}'_{\Delta t}
&\textrm{if } \mathbf{X}_{\Delta t} = \mathbf{X}
\numberthis 
\end{align*}

Finally, we write down the bias-corrected expression for the objective function $\mathcal{L}'$ :
\begin{align*}
\label{eq:correction}
\mathcal{L}'&= \frac{c}{N} \sum \limits_{n=1}^N e^{\beta V^n} \ln Q(\mathbf{X^n}| \bm{\chi^n}) \\
&  \quad - \frac{c}{N} \sum \limits_{n=1}^N  e^{\beta \frac{V^{n+k}+V^n}{2}} \ln Q(\mathbf{X^n}| \bm{\chi^n}) \\ 
&  \quad + \frac{c}{N} \sum \limits_{n=1}^N  e^{\beta \frac{V^{n+k} +V^n}{2}}  \ln Q(\mathbf{X^{n+k}}| \bm{\chi^n}) \numberthis 
\end{align*}
where $c$ is an irrelevant constant independent of the neural network parameters. 

The last term in Eq. \ref{eq:correction} is similar to Eq. \ref{eq:PIB_vb3}, which only considers the correction for sampling points and not their dynamical propagation. The summation of the first two terms can be interpreted as the correction for the propagator. To gain some intuition into these, let's consider the effect of the biasing potential on the dynamics. It pushes the system from high bias region to low bias region, thereby spuriously increasing the conditional probability for being found in low bias region, given that earlier the system was in high bias region. We thus want to reweight out this effect and only learn from the dynamics induced by the original potential, which is what Eq. \ref{eq:correction} achieves. If the trajectory goes from a state $\mathbf{X^{n}}$ with higher bias $V^{n}$ to a state $\mathbf{X^{n+k}}$ with lower bias $V^{n+k}$, the summation of the first two terms in Eq. \ref{eq:correction} will be positive. Note that in Eq. \ref{eq:correction} the first two terms together in this regime encourage the decoder to generate features close to $\mathbf{X^{n}}$ while the third term favors the decoder to generate features closer to $\mathbf{X^{n+k}}$. This will negate the spurious enhancement of probability of $\mathbf{X^{n+k}}$ due to the biasing profile. Similar argument applies for the case $V^{n+k}>V^{n}$. We conclude this section by highlighting that is a rough approximation valid for short $\Delta t$, and it can not completely correct the effect of bias on dynamics. In the following examples we will discuss the advantages and limitations of this new formula, and also analyze how valid was the intuitive approximation Eq. \ref{eq:approximation} made in Ref. \onlinecite{prave}.

\begin{figure}
     \centering
     \begin{subfigure}[b]{0.23\textwidth}
         \includegraphics[width=\textwidth]{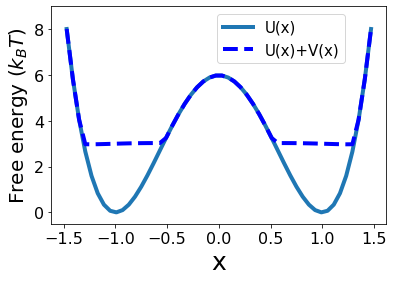}
         \caption{}
         \label{fig:fes_x}
     \end{subfigure}
     \hfill
     \begin{subfigure}[b]{0.23\textwidth}
         \includegraphics[width=\textwidth]{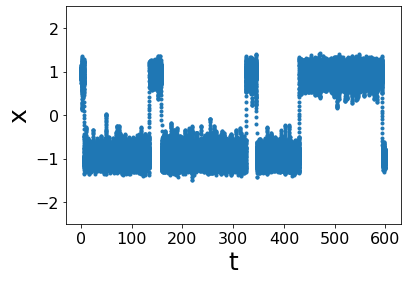}
         \caption{}
         \label{fig:traj_x}
     \end{subfigure}
     \begin{subfigure}[b]{0.23\textwidth}
         \includegraphics[width=\textwidth]{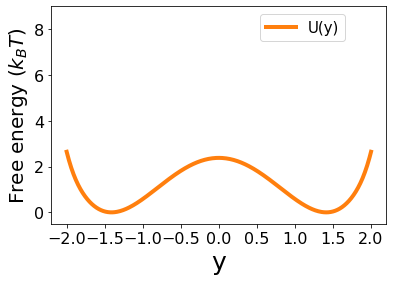}
         \caption{}
         \label{fig:fes_y}
     \end{subfigure}
        \hfill
     \begin{subfigure}[b]{0.23\textwidth}
         \includegraphics[width=\textwidth]{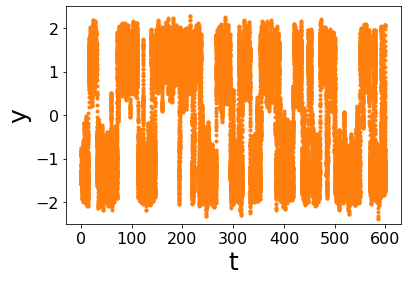}
         \caption{}
         \label{fig:traj_y}
     \end{subfigure}
        \hfill
     \begin{subfigure}[b]{0.23\textwidth}
         \includegraphics[width=\textwidth]{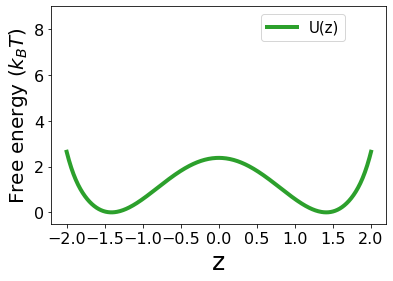}
         \caption{}
         \label{fig:fes_z}
     \end{subfigure}
          \hfill
     \begin{subfigure}[b]{0.23\textwidth}
         \includegraphics[width=\textwidth]{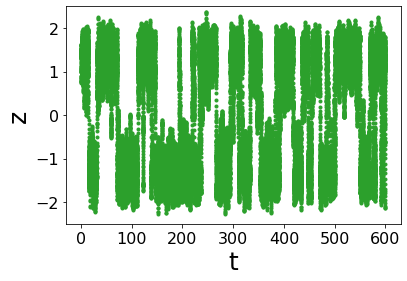}
         \caption{}
         \label{fig:traj_z}
     \end{subfigure}
        \caption{Free energy profile along each degree of freedom of the model system ((\subref{fig:fes_x}), (\subref{fig:fes_y}) and (\subref{fig:fes_z}) respectively) and the corresponding  trajectories from Langevin dynamics ((\subref{fig:traj_x}), ((\subref{fig:traj_y})) and (\subref{fig:traj_z}) respectively). In (\subref{fig:traj_x}) the sum of underlying and bias potentials are indicated with a dashed line.}
        
        \label{fig:system}
\end{figure}

\section{Results}
\label{sec:Results}
We now demonstrate the nature and effect of the corrections derived in the previous section through a simple illustrative numerical example with exact results pertaining to the true reaction coordinate and its free energy. More specifically, we generated an unbiased trajectory on a model system and apply RAVE iterations to it. This system is simple enough that we could generate a long enough unbiased trajectory with sufficient sampling of different metastable states. RAVE calculations on this perfect unbiased trajectory serve as our benchmark. In parallel, we add a biasing potential along the relevant slow mode of the system to generate a biased trajectory. Both of these trajectories are then subjected to different treatments as proposed in Sec. \ref{sec:theory}, and as explained in the remainder of this section, in order to judge how much of a difference our corrections really make.

\subsection{System set-up}
\label{sec:systemsetup}

 The model system in this work has three degrees of freedom and a governing potential energy $U$ given by:
\begin{equation}
    U(x, y, z)= 6(x^2-1)+0.0375[(y-z)^2-8]^2 + 45(y+z)^2
    \label{eq:potential}
\end{equation}

Fig. \ref{fig:system} shows the free energy profile along each degree of freedom and the corresponding trajectory from evolving Langevin dynamics\cite{Bussi_Parrinello_thermostat} with an integration time step of 0.01 units at temperature $k_BT = 1$. In agreement with the significantly higher energy barrier along $x$, the transitions along $x$ happen much less frequently and it represents accurately the dominant slow mode in this system.

In this work we want to focus on the difference between optimizing Eq. \ref{eq:PIB_vb3} and Eq. \ref{eq:correction}. As such we perform the biased MD by using the slow mode $x$ as the reaction coordinate and constructing the bias potential on the basis of its analytical free energy. The bias potential was constructed with a maximum bias of $3kT$. The sum of the potential energy of the system and the bias potential is shown as a dashed line in Fig. \ref{fig:system} (\subref{fig:fes_x}). 

\subsection{RAVE set-up}
\label{sec:ravesetup}
The input to RAVE comprised the three-dimensional time-series of $\{x,y,z \}$ as obtained from Langevin dynamics. Whitening procedure from Ref. \cite{whitening} was used to reduce artifacts from high variance. The RC or PIB was learnt as a linear combination of $\{x,y,z \}$, thus characterized by corresponding three weights. Due to the possible non-convex and inherent stochastic nature of the optimization, there is no guarantee that a single training run gives us the most optimal RC. Indeed, here we trained 16 neural networks with randomly initialized weights, and found significant run to run variation in the output weights for $\{x,y,z \}$. Out of the 16 choices we selected the run that gave the smallest loss function.
In practice we believe such a high number of trial runs is not needed, and a smaller number might very well be sufficient. 

\subsection{Calculated RC under different time-delays and approximation schemes}
\label{sec:time delay}
\begin{figure}[H]
     \centering
     \begin{subfigure}{0.4\textwidth}
         \includegraphics[width=0.8\textwidth]{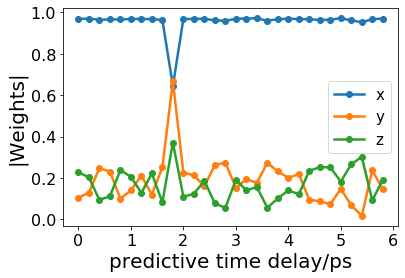}
         \caption{}
         \label{fig:unbiased}
     \end{subfigure}
     \hfill
     \begin{subfigure}{0.4\textwidth}
         \includegraphics[width=0.8\textwidth]{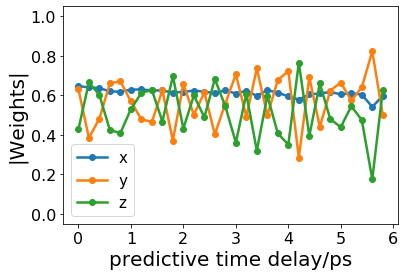}
         \caption{}
         \label{fig:nocorrection}
     \end{subfigure}
     \begin{subfigure}{0.4\textwidth}
         \includegraphics[width=0.8\textwidth]{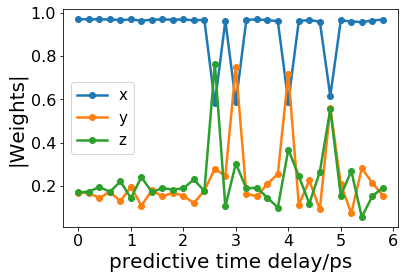}
         \caption{}
         \label{fig:correction1}
     \end{subfigure}
     \begin{subfigure}{0.4\textwidth}
         \includegraphics[width=0.8\textwidth]{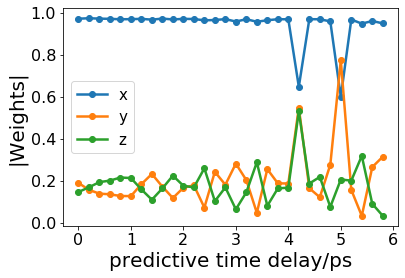}
         \caption{}
         \label{fig:correction2}
     \end{subfigure}
        \caption{Reaction coordinates learned by RAVE as functions of predictive time delay. (\subref{fig:unbiased}) RC learned from an unbiased trajectory using Eq. \ref{eq:PIB_vb3} setting $V^n=0$. (\subref{fig:nocorrection}) RC learned from a biased trajectory using Eq. \ref{eq:PIB_vb3} with $V^n=0$. (\subref{fig:correction1}) RC learned from a biased trajectory with Eq. \ref{eq:PIB_vb3} taking the bias $V^n=$ into account for reweighting the stationary probability, but not the propagator. (\subref{fig:correction2}) RC learned from a biased trajectory correcting for the effect of bias on the stationary density as well as the propagator by making use of Eq. \ref{eq:correction}. }
        \label{fig:time_delay}
\end{figure}

We use the unbiased and the biased trajectories as input in RAVE with different values of the predictive time-delay $\Delta t$. Fig. \ref{fig:time_delay} shows the weights of $x$,$y$ and $z$ in the RC as obtained for these different set-ups. For the long well-sampled unbiased trajectory, we use Eq. \ref{eq:PIB_vb3} setting $V^n=0$ (Fig. \ref{fig:time_delay} (\subref{fig:unbiased})). This serves as a benchmark. For the biased trajectory as well, we first use Eq. \ref{eq:PIB_vb3} setting $V^n=0$ (Fig. \ref{fig:time_delay} (\subref{fig:nocorrection})). This illustrates the effect of biasing on the different modes. Next, for the biased trajectory, we use Eq. \ref{eq:PIB_vb3} taking the bias into account for reweighting the stationary probability, but not the propagator (Fig. \ref{fig:time_delay} (\subref{fig:correction1})). Finally, we use the biased trajectory correcting for the effect of bias on the stationary density as well as the propagator by making use of Eq. \ref{eq:correction} (Fig. \ref{fig:time_delay} (\subref{fig:correction2})). 

The RAVE solution for unbiased trajectory shown in Fig. \ref{fig:time_delay} (a) assigns the highest weight to $x$, which is in agreement with our expectation that $x$ is the slowest mode. The  weights for $y$ and $z$ always have different signs as they are anticorrelated according to potential energy $U$ in Eq. \ref{eq:potential}. Here we just want to compare their relative importance so only the absolute values of the weights are shown all throughout Fig. \ref{fig:time_delay}. Fig. \ref{fig:time_delay} (\subref{fig:nocorrection}) shows results after treating the biased trajectory without taking the bias into account i.e. through the use of Eq. \ref{eq:PIB_vb3} with $V^n=0$. Due to biasing along $x$ and subsequent enhancement in fluctuations in this direction, now all 3 modes $x$, $y$ and $z$ are comparable in their timescales, and without any reweighting RAVE assigns equal weights to the three degrees of freedom. This illustrates why reweighting out effects of bias is crucial in RAVE, and arguably in general when using inputs from biased simulations as training data in machine learning. In Figs.  \ref{fig:time_delay} (\subref{fig:correction1}) and (\subref{fig:correction2}) we use Eq. \ref{eq:PIB_vb3} and Eq. \ref{eq:correction} respectively to correct for the effect of bias on the stationary probability and both stationary probability/propagator. Both Eqs. \ref{eq:PIB_vb3} and Eq. \ref{eq:correction} in Figs.  \ref{fig:time_delay} (c) and (d) respectively give a remarkably similar profile to that obtained from the long benchmark unbiased simulation used as input in RAVE. Thus, at least for this model system, it appears that our intuitive approximation of Eq. \ref{eq:approximation} made in Ref. \onlinecite{prave} was reasonable. It does however appear that using the correction Eq. \ref{eq:correction} developed in this work leads to a more robust solution across different predictive time delays.

\section{Conclusion}
In this work, we have revisited our recent iterative machine learning--molecular dynamics method RAVE \cite{rave,multirave,prave}. Specifically, we first discuss the role of predictive time delay in RAVE demonstrating why its specific value is not relevant as long as a small non-zero value is taken. Secondly, we introduced a correction for the objective function in RAVE that corrects the effect of biasing potential on the dynamical propagator of the system. Our work is grounded in the master equation framework for the dynamics of the system in the true high-dimensional space. We prove that the RC learned from RAVE should not depend on the choice of time delay, as long as time delay is small enough that the Master Equation formalism of Eq. \ref{eq:mastereq2} is valid. This explains why in our previous work, the RC converged quickly as a function of predictive time delay\cite{prave}. Also, by introducing the correction for the transition probability in biased MD, we derive a new objective function, which not only reweights the static distribution as we did in our previous work, but also gives a better estimation of the transition probabilities. We find that apart from reducing the number of outlier solutions, our correction does not significantly improve upon the intuitive approximation introduced in Ref. \onlinecite{prave}. It remains to be seen if this holds true in more complex systems, but given that our correction Eq. \ref{eq:correction} involves no computational overhead relative to Ref. \onlinecite{prave}, we recommend its use as a default. In future work we will construct even more accurate approximations\cite{reweight_dynamic} to account for the effect of bias on transition probabilities, as well as apply this framework to different practical applications. \newline

\textbf{Acknowledgements\newline }
The authors thank Sun-Ting Tsai for sharing the code implementing Langevin dynamics and Pavan Ravindra for proof-reading the manuscript. The authors thank Deepthought2, MARCC, and XSEDE for providing computational resources used in this work. YW would like to thank NCI-UMD Partnership for Integrative Cancer Research for financial support. Acknowledgment is made to the Donors of the American Chemical Society Petroleum Research Fund for partial support of this research (PRF 60512-DNI6). \newline

\textbf{References}

	\end{document}